\documentclass[%
preprint,
nofootinbib,
 amsmath,amssymb,
 aps,
showkeys
]{revtex4-2}
\usepackage{adjustbox}
\usepackage{graphicx}% Include figure files
\usepackage{dcolumn}% Align table columns on decimal point
\usepackage{bm}% bold math
\usepackage{soul} % to cross a text; command \st{....}
\usepackage{color}% print color text
\definecolor{vs}{rgb}{0.1,0.4,0.1}                  % dark green
\newcommand{\del}[1]{}                             % to accept all deletions comment \newcommand{\del}[1]{\textcolor{vs}{\st{#1}}} with % and uncomment this line
\usepackage{hyperref}% add hypertext capabilities
\usepackage{verbatim}
\newcommand\wordcount{\verbatiminput{\jobname.sum}}
\begin{document}

%\preprint{}

\vspace*{5cm}
\title{Two tractable models of non-stationary light scattering by subwavelength particles and their application to Fano resonances}% Force line breaks with \\

\author{Michael I. Tribelsky}
\email{mitribel@gmail.com}
 \affiliation{M. V. Lomonosov Moscow State University, Faculty of Physics, Moscow, 119991, Russia}
 \altaffiliation[Also at ]{National Research Nuclear University MEPhI (Moscow Engineering Physics Institute), Moscow, 115409, Russia}%Lines break automatically or can be forced with \\
\author{Andrey E. Miroshnichenko}%
 \email{andrey.miroshnichenko@unsw.edu.au}
\affiliation{%
 University of New South Wales Canberra, School of Information and Information Technology, ACT, 2600, Australia
}%

\date{\today}% It is always \today, today,
	
\begin{abstract} 
We introduce two tractable analytical models to describe dynamic effects at resonant light scattering by subwavelength particles. One of them is based on generalization of the temporal coupled-mode theory, and the other employs the normal mode approach. We show that sharp variations in the envelope of the incident pulse may initiate unusual, counterintuitive dynamics of the scattering associated with interference of modes with fast and slow relaxation. To exhibit the power of the models, we apply them to explain the dynamic light scattering of a square-envelope pulse by an infinite circular cylinder made of $GaP$, when the pulse carrier frequency lies in the vicinity of the destructive interference at the Fano resonances. We observe and explain intensive sharp spikes in scattering cross section just behind the leading and trailing edges of the incident pulse. The latter occurs when the incident pulse is over and is explained by the electromagnetic energy released in the particle at the previous scattering stages. The accuracy of the models is checked against their comparison with results of the direct numerical integration of the complete set of Maxwell's equations and occurs very high. The models' advantages and disadvantages are revealed, and the ways to apply them to other types of dynamic resonant scattering are discussed.
%
%\del{We study numerically and analytically effects of resonant light scattering by subwavelength high-index particles with weak dissipation in the vicinity of the destructive interference at Fano resonances.  %[Phys. Rev. A, vol. 100, 053824 (2019)].
%We show that sharp variations in the envelope of the incident pulse may initiate unusual, counterintuitive dynamics of the scattering associated with interference of modes with fast and slow relaxation. In particular, we observe and explain intensive sharp spikes in scattering cross section just behind the leading and trailing edges of the incident pulse. The latter occurs when the incident pulse is over and is explained by the electromagnetic energy released in the particle at the previous scattering stages. To mimic the numerical results, we develop two tractable analytical models. Both reproduce with high accuracy all the dynamic effects of the numerics. The models allow us to reveal the physical grounds for the spikes explained by the violation of balance between the resonant and background partitions during the transient. Besides, we compare the models with each other and reveal their mutual advantages and disadvantages.}
\end{abstract}


\maketitle

\section{Introduction}

High-$Q$ resonances are of utmost importance in a wide diversity of problems \cite{Huang2021}. {It is explained by the fact that to obtain strong resonant effects, the corresponding resonance should have a high amplitude, and hence a high $Q$-factor,} {at least if a spatially-bounded system is a concern.} However, the characteristic relaxation time for a resonance is inversely proportional to its $Q$-factor. Thus, the price one must pay for making use of high-$Q$ resonances is long-lasting transient effects. On the other hand, the frontier of modern photonics moves toward short and ultrashort pulses. Nowadays, Currently, these two factors together make a typical situation where the duration of the laser pulse becomes comparable or even shorter than the relaxation time of the resonance effects initiated by this pulse.

{ Though in quantum spectroscopy, it is well-known that non-steady resonant scattering may qualitatively differ from its steady-state realizations, see, e.g., Ref.~\cite{kaldun2016observing}, the corresponding studies in light scattering by subwavelength particles have begun only recently~\cite{Tribelsky_Mirosh:PRA_2019,svyakhovskiy2019anapole,avalos2020temporal}. As it could be expected, these studies also reveal qualitatively new effects, which do not exist at the steady-state scattering. However, for the time being, analytical descriptions of the resonant light scattering by particles are still based on the solutions of Maxwell's equations describing the steady-state scattering, where the processes of gradual "swinging" of resonant modes are not taken into account. }

{Thus, at the moment, the only possible theoretical description of transient effects at the resonant light scattering by subwavelength particles is made with the help of direct numerical integration of the complete set of Maxwell's equations. Due to the existence of various numerical methods and many pieces of software, both free and commercial, created to perform this integration, it has become a more or less routine procedure. However, to find in the range of the problem parameters windows, where the desired effect is the most pronounced, the dependence of the scattering on these parameters in a wide domain of their variations is required.  To this end, numerical methods are not appropriate, and analytically tractable models are required. To the best of our knowledge, for the time being, none of them exists.}

Here, we present two such models, apply them to describe non-steady Fano resonances, and compare the results with direct numerical integration of the complete set of Maxwell's equations. %\del{discussed in Ref.[1]}.
The comparison indicates the high accuracy of both models and reveals their mutual advantages and disadvantages. The models unveil the physical grounds for the counterintuitive spikes in the scattered radiation observed behind the leading and trailing edges of the incident pulse.

{The first model is based on the temporal coupled-mode theory (TCMT)~\cite{louisell1960coupled} generalized to applications to essentially
non-steady scattering. The second model mimics non-steady resonant vibrations by a superposition of dynamics of driven harmonic oscillators (HO). The latter approach looks similar to the harmonic inversion, see, e.g., Ref.~\cite{Mandelshtam:JCP:1997,barone1989segmented,roessling2015finite}.
However, in our case, the mode selection for the approximation and, most importantly, the choice of the values of the model parameters are based on the system in question's physical properties. Thus, it reveals the role of different excitations in the system dynamic and sheds light on the physical nature of the system as a whole. Besides, this makes it possible to reduce the number of modes to be studied just to a few with non-trivial dynamics.} % As a result, our method occurs considerably simpler than others and often may provide closed-form analytical solutions describing complicated dynamics with high accuracy.
%This paper presents a general approach to the theoretical description of non-steady resonant light scattering by subwavelength particles based on the extension of the Mie solution to this case. We apply it to inspect the light scattering by a cylinder. We show that non-steady resonant scattering of light may exhibit unexpected new effects, and the scattering field patterns may have very little in common with the one for the steady-state scattering.
{ In addition to the purely academic interest, the results obtained may be employed to design a new generation of fast, multifunctional passive nanodevices whose properties vary quantitatively with a variation of the duration of the incident pulse.}

%{To illustrate the power of the approaches they are applied to describe and explain the results of Ref.~\cite{Tribelsky_Mirosh:PRA_2019},
%where a dynamic scattering of a rectangular pulse by a cylinder %at the pulse carrier frequency close to destructive Fano resonances ones
%was studied by direct numerical integration of the complete set of Maxwell's equations. The comparison of the dynamic given by both
%approaches with the numerics exhibits their high accuracy.}

The paper has the following structure: In Sec. 2 the problem formulation is presented. Sec. 3 is devoted to the problem analysis and discussion of the obtained results. In Sec. 4 we formulate conclusions. Cumbersome details and expressions and several plots illustrating the developed approach are moved to Appendix.

\section{Problem formulation}
\subsection{Preliminary}

The simplest exactly solvable light scattering problems correspond to the steady-state scattering of a linearly polarized plane wave by a homogeneous sphere (the Mie solution) or infinite circular cylinder. In these solutions, the scattered field is presented as an infinite series of partial waves (dipole, quadrupole, etc.), also called multipoles~\cite{Bohren::1998}. Here, we generalize this approach to dynamic light scattering, making it possible to study quite intricate transient features.

One of the most typical resonant responses in light scattering by finite obstacles is the Fano resonance~\cite{Miroshnichenko:2010ewa}. A characterizing it asymmetric lineshape is explained by either constructive (the maximal scattering) or destructive (the minimal scattering) interference occurring close to each other in the frequency domain. The interfering parties are the so-called resonant (narrow line) and background (broad line) partitions of the same multipole. The narrow-line and broad-line partitions correspond to excitations with slow and fast relaxation time in the time-domain, respectively.

Naturally, the procedure of splitting a single partial wave into the two partitions is not unique. Accordingly, there are two main equivalent approaches to it. In the first approach, a partial wave is presented as a sum of the radiation of the conventional electric and toroidal multipoles~\cite{Miroshnichenko2015}. However, in what follows, it will be more convenient for us to employ another approach~\cite{Tribelsky_Mirosh_2016}. In this approach, the resonant partition is associated with the corresponding electromagnetic mode excited in the bulk of the particle (volume polariton). In contrast, the background partition corresponds to the radiation of the surface current induced by the same incident wave, scattered by the particle with the same geometrical shape but made of a hypothetical material called the perfect electric conductor (PEC).

{At a point of the destructive interference, the two partitions cancel each other. As a result, the contribution of the corresponding multipole to the stationary scattering is totally suppressed.} For a subwavelength particle, when just a few first multipoles produce the overwhelming contribution to the overall scattering, the suppression even of one of them may reduce the scattering cross section dramatically~\cite{Tribelsky_Mirosh_2016}. However, since, as has been mentioned above, the resonant and background partitions are characterized by substantially different relaxation times, it is evident that during transient regimes, the mutual cancelation does not occur. Therefore, the violation of the destructive interference conditions must give rise to a considerable increase in the scattering intensity during the transient and other unusual effects. Some of them have already been discussed in our previous publications~\cite{Tribelsky_Mirosh:PRA_2019,svyakhovskiy2019anapole}.

We stress that though this increase in the scattering intensity looks similar to the well-known overshoot effect when a driven high-$Q$ oscillator exhibits oscillatory relaxation to the steady-state, the physical grounds for the former is entirely different: If the overshoot is related to vibrations of a {\it single\/} oscillator, the discussed effect is explained by a superposition of {\it two different\/} oscillations. This distinction in the physical nature of the two cases gives rise to the corresponding difference in their manifestations. In particular, {in contrast to the overshoot, the transient at the point of the destructive interference corresponds to relaxation to the steady-state with {\it zero\/} amplitude of the given multipole. Therefore, for a small particle with just a few dominant multipoles, the amplitude of the spike in the overall scattering may be {\it in orders of magnitude\/} larger than that exhibited by the overshoot.}

Thus, the dynamic resonant light scattering is a new, rich, and practically untouched subfield. Plenty of exciting effects hidden there are still undiscovered. In the present paper, we continue to explore this appealing topic.

To understand typical main features of the phenomenon, we consider the simplest problem formulation {corresponding to the one employed in Ref.~\cite{Tribelsky_Mirosh:PRA_2019} for the computer simulation}, namely the scattering of a square incident pulse with duration $\tau$ (the amplitudes of the electromagnetic fields \mbox{$A(t)=A_0= const \neq 0$} inside the pulse and zero \mbox{outside it}), carrier frequency $\omega$, and temporal dependence of the fields $\sim\exp(-i\omega t)$ by an infinite circular {high-index} cylinder with the base radius $R$ and complex refractive index $m=n+i\kappa$ {$(n\gg 1,\;\kappa \ll 1)$}. The refractive index of the surrounding cylinder medium equals unity. The cylinder is nonmagnetic, so its permeability $\mu = 1$. For further simplification, just the TE polarization and normal incidence are considered, see below Fig.~\ref{Efficiensies}(a).

\subsection{Instantaneous scattering cross section}
In the conventional steady-state case, the scattering is quantitatively described by the cross section $C_{\rm sca}$ calculated per unit of
length of the cylinder. $C_{\rm sca}$ is defined as the ratio of the integral power flux through a closed remote surface surrounding the
scatterer to the intensity of the incident light. At non-steady scattering, both these quantities are time-dependent, and their ratio is
not a constant anymore. Moreover, the ratio depends on the shape and position of the surface used to calculate the flux since the speed of
light is finite.

To describe non-steady scattering, we introduce the \emph{instantaneous\/} scattering cross section $C_{\rm sca}(t)$ as the ratio of the instantaneous value of the power flux through a cylindrical surface, coaxial with the scattering cylinder and lying in the far wave zone, calculated per a unit of length of the cylinder, to the constant intensity of the square incident pulse
$I_0$~\cite{Tribelsky_Mirosh:PRA_2019}.{ In a more general case, when the pulse envelope has a time-dependent shape $I(t)$, the scattered
flux may be normalized over, e.g., the maximal value of $I(t)$, i.e., $I_0 = \mathop{\rm Max}\limits_t\{I(t)\}$.

The corresponding dimensionless scattering efficiency, $Q_{\rm sca}(t)$ is connected with $C_{\rm sca}(t)$ by the usual relation $Q_{\rm
sca}(t) = C_{\rm sca}(t)/(2R)$, where $R$ is the radius of the base of the cylinder. We also do not perform the {time} averaging of the
Poynting vector {over the period of oscillations.}} %Generalization of $C_{\rm sca}(t)$ to a spherical particle is straightforward.

The exact solution describing the steady-state scattering is built as an infinite series of partial waves (multipoles). For the problem in
question, the complex amplitudes of the multipoles (scattering coefficients) associated with the outgoing partial waves $a_\ell$ and the
field within the cylinder $d_\ell$ are given by the well-known expressions presented in Appendix Eqs. \eqref{eq:a_ell}--\eqref{eq:d_ell}, see also, e.g., \cite{Bohren::1998}. Here $\ell$ designates the multipole order (dipole, quadrupole, etc.). {Note that \mbox{$a_\ell = a_{-\ell}$,} \mbox{$d_\ell = d_{-\ell}$} ~\cite{Bohren::1998}.}
%
%\begin{eqnarray}
% a_\ell &=& \frac{mJ_\ell(mx)J^{\prime}_\ell(x) - J_\ell(x)J_\ell'(mx)}{mJ_\ell(mx)H^{(1)\prime}_\ell(x) - H^{(1)}_\ell(x)J_\ell'(mx)},
%\label{eq:a_ell} \\
% d_\ell &=& \frac{{2i}/{(\pi x)}}{mJ_\ell(mx)H^{(1)\prime}_\ell(x) - H^{(1)}_\ell(x)J_\ell'(mx)}\label{eq:d_ell}\;,
%\end{eqnarray}
 %({$\ell=0$ and $\ell=1$ corresponds to dipoles of orthogonal
%polarizations;} $\ell = 2$ denotes a quadrupole, etc.);

For the given problem, the scattering coefficients satisfy the identity~\cite{Tribelsky_Mirosh_2016}:
\begin{equation}\label{eq:al-dl}
a_\ell \equiv a_\ell^{(\rm PEC)} - \frac{J^\prime_\ell(mx)}{H^{(1)\prime}_\ell(x)}d_\ell;\; a_\ell^{(\rm PEC)} \equiv
\frac{J^\prime_\ell(x)}{H^{(1)\prime}_\ell(x)},
\end{equation}
where $x = kR$ stands for the size parameter; \mbox{$k=\omega/c$;} $c$ is the speed of light in a vacuum; $J_\ell(z)$ and $H^{(1)}_\ell(z)$
designate the Bessel and Hankel functions, respectively; prime denotes the derivative over the entire argument of a function; and $
a_\ell^{(\rm PEC)}$ is the scattering coefficient of the same cylinder made of the perfect electric conductor.

Routine calculations result in the following expressions~\cite{Tribelsky_Mirosh:PRA_2019}:
\begin{eqnarray}
\!\!\!\!\!\!\!\!& & Q_{\rm sca}=Q_{\rm sca}^{(0)} + Q_{\rm sca}^{(\rm osc)} = \sum_{\ell=-\infty}^{\infty}\left\{Q^{(0)}_{{\rm
sca}\,(\ell)}+ Q_{{\rm sca}\,(\ell)}^{(\rm osc)}\right\},%\; Q_{{\rm sca}\,(\ell)} = Q_{{\rm sca}\,(\ell)}^{(0)} + Q_{{\rm sca}\,(\ell)}^{(\rm osc)},
\label{eq:Qsca} \\
\!\!\!\!\!\!\!\!& & Q_{{\rm sca}\,(\ell)}^{(0)}\! =\! \frac{2}{x}|a_\ell|^2;\; Q_{{\rm sca}\,(\ell)}^{(\rm osc)}\! = \!
-\frac{i}{x}\!\left[a_\ell^2 e^{2i(kr-\omega t)}\!-\!c.c.\right]\!\!, \label{eq:Q_through_a}
\end{eqnarray}
Here $Q_{\rm sca}^{(0)}$ is the conventional scattering efficiency, while $Q_{\rm sca}^{(\rm osc)}$ is an additional rapidly oscillating in
time and space term with zero average {and $r$ lies in the far-wave zone ($kr \gg 1$)}.

\subsection{Fano resonances}
{The Fano resonances~\cite{Fano:NC:1935,Fano:PR:1961,Miroshnichenko:2010ewa} are a good example demonstrating unusual, counterintuitive
effects in transient processes of resonant light scattering. For the steady-state scattering, their detailed discussion is presented, e.g., in Ref.~\cite{Tribelsky_Mirosh_2016}. Though in {that} paper, a spherical particle is considered, generalization to the case of a cylinder is a straightforward matter.

As it has been mentioned above, a key point of the Fano resonances is a presentation of the scattered wave as a sum of two partitions: resonant
and background. In the proximity of the resonance the amplitude and phase of the former have a sharp \mbox{$\omega$-dependence,} while for the latter its $\omega$-dependence is weak. An important conclusion following from the results of {Ref.~\cite{Tribelsky_Mirosh_2016,Rybin2013a}} is that the splitting of $a_\ell$ into the two terms, given by Eq.~\eqref{eq:al-dl}, actually, is the singling out the background ($a_\ell^{(\rm PEC)}$) and resonant ($- \frac{J^\prime_\ell(mx)}{H^{(1)\prime}_\ell(x)}d_\ell$) partitions.

{Our goal is to recover the full time-dependence $Q_{\rm sca}(t)$. For high-$Q$ resonances, which we are interested in, the
characteristic time scales of the transients should be large relative to the period of the field oscillations $2\pi/\omega$. Then, a
quasi-steady approximation may be employed. It implies the same structure of the solution like that for steady-state scattering. However,
now the scattering coefficients are regarded as \emph{slowly-varying\/} functions of time.}
Let us apply this assumption to the TCMT and HO.%the two approaches mentioned above to recover these functions.}% {It is essential to stress that, though the approaches are employed to describe this specific problem of resonant light scattering, they may be readily extended to inspect a broad class of non-steady resonant phenomena.}

%To do that, we are going to employ two approaches described above, namely TCMT and forced coupled harmonic oscillators, where we replace
%in Eq.~\eqref{eq:al-dl} \mbox{$a_\ell \rightarrow a_\ell(t)$}, \mbox{$d_\ell \rightarrow d_\ell(t)$}.

%\section{Models}
%a{\bf AEM started here}

\subsection{Temporal coupled-mode theory}
%\subsubsection{{Problem formulation}}
%
%In this case, the internal degrees of freedom are modeled through the eigenmode oscillations. {For problems with real coefficients, each
%eigenmode has a complex conjugated counterpart. Their dynamics may be presented as co- and counterclockwise rotations in the complex plane.
%The main advantage of the TCMT is that if the coupling between the rotating in opposite directions modes is weak, the two sets may be
%approximately regarded as independent of each other. This reduces by half the order of the corresponding differential
%equation(s)~\cite{louisell1960coupled}.}
%
In the specified case, %of light scattering
the TCMT equations read as follows, see, e.g., Ref.~\cite{Fan:TCM_2010}:

\begin{eqnarray}
& & \frac{\mathrm{d} p(t)}{\mathrm{d} t} =-\left({i} \omega_{0}-\gamma\right) p(t)+\kappa s^{+}(t) \label{eq:TCMT_p} \\
& & s^{-}(t) =B s^{+}(t)+\zeta p(t) \label{eq:TCMT_s}
\end{eqnarray}
Here $s^+(t)$ and $s^-(t)$ are the amplitudes of the incoming (converging) and outgoing (diverging) cylindrical waves, respectively; $B$
is the background reflection coefficient; $\kappa$ and $\zeta$ are coupling constants; $p(t)$ describes the internal resonant mode
excitation; and $\hat{\omega}_{0}\equiv \omega_{0}+i\gamma$ is the nearest pole of the scattering coefficients in the plane of complex
$\hat{\omega}$. {It is important to stress that for the selected temporal dependence $\sim \exp(-i\omega t)$ decaying modes must have
\mbox{$\gamma <0$.} Then, \emph{the corresponding poles are situated in the lower semiplane.} }

The analysis of Eqs.~\eqref{eq:TCMT_p}, \eqref{eq:TCMT_s} performed in Ref.~\cite{Fan:TCM_2010} for the steady-state scattering indicates
that \mbox{$\kappa =\zeta =\sqrt{2|\gamma|}\exp(i\theta)$}; $B=\exp(i\phi)$ and {$\theta=(\phi + \pi)/2 + n\pi$,} where $n$ is an arbitrary
integer. %, which may be put to zero (see below).
In this case the steady-state scattering coefficient $a_\ell^{\rm(TCMT)}$ is defined as $[s^{-}(t)-s^{+}(t)]/({2s^{+}(t)})$. Eventually, it
gives rise to %the following expression~\cite{Fan:TCM_2010}:
a certain expression for $a_{\ell}^{\rm(TCMT)}$, where phase $\phi$ remains undefined yet.
%
%\begin{eqnarray}\label{eq:a_ss}
%\begin{aligned} a_{\ell}^{\rm(TCMT)} %&=\frac{1}{2}(r-1) \\ &
%=\frac{1}{2} \frac{{i}\left(\omega_{0}-\omega\right) ({e}^{{i} \phi}-1 )-\gamma\left(1+{e}^{{i}
%\phi}\right)}{{i}\left(\omega_{0}-\omega\right)+\gamma}. \end{aligned}
%\end{eqnarray}
%%
%Here, the value of $\phi$ is undefined yet.
The authors of Ref.~\cite{Fan:TCM_2010} fix it by fitting the profile $|a_{\ell}^{\rm(TCMT)}(\omega)|^2$ to $|a_{\ell}(\omega)|^2$ obtained
from the steady-state exact solution. {However, any fitting procedure is ambiguous since its results depend on the fitting window's size.

Meanwhile, there are other ways to fix $\phi$, free from this disadvantage. In this paper we fix $\phi$ from the condition
\mbox{$|a_{\ell}^{\rm(TCMT)}(\omega)|=|a_{\ell}(\omega)|$} at the carrier frequency of the incident pulse. This gives rise to a quadratic
equation, whose solution results in two values of $\phi$ in the non-trivial domain \mbox{$-\pi \leq \phi \leq \pi$}. The final choice is
made based on the better overall coincidence of the two profiles. Such a choice is a straightforward matter, see Appendix.}

The next difficulty is that the employed expression for $a_\ell^{\rm (TCMT)}$ is valid for the steady-state scattering solely. The latter
is evident if we consider, e.g., the case when the incident pulse is already over, i.e., $s^+(t)=0$. At the same time, the particle still radiates
the accumulated electromagnetic energy, so that $s^-(t)\neq 0$. In this case, the discussed expression diverges. %The physical grounds for the divergence are the same as those discussed in paragraph \emph{Instantaneous scattering cross section}.
To avoid this difficulty, we have to redefine $a_\ell^{\rm (TCMT)}$. For the considered square pulse, it may be done as follows:
$a_\ell^{\rm (TCMT)}(t)=\frac{1}{2A_{0(\ell)}\exp(-i\omega t)}[s^{-}(t)-s^{+}(t)]$, where $A_{0(\ell)}$ is a constant amplitude of a converging incident cylindrical wave, corresponding to a given multipolarity. This definition coincides with the above one for the
steady-state scattering but remains finite at $s^+(t)=0$. In an arbitrary pulse shape, the role of $A_{0(\ell)}$ may play the corresponding
maximal value.

Regarding $Q_{\rm sca}(t)$, in the discussed quasi-steady approximation the solution retains the same structure as that for the
steady-sate. Then, Eqs.~\eqref{eq:Qsca}, \eqref{eq:Q_through_a} still remain valid but the replacement \mbox{$a_\ell \rightarrow a_\ell^{\rm
(TCMT)}(t)$ is required.} As for the dependence $a_\ell^{\rm
(TCMT)}(t)$, it is readily obtained by integration of Eq.~\eqref{eq:TCMT_p} with the initial condition $p(0)=0$:

\begin{eqnarray}
& & a_\ell^{\rm (TCMT)}(t) = \label{eq:a_l^TCMT_t<tau} \frac{i \gamma \left[1-e^{i \phi } \left(2 e^{it (\omega -\omega_0)+\gamma t}-1\right)\right]+\left(e^{i \phi }-1\right) (\omega
-\omega_0)}{2 (\omega -\omega_0 - i \gamma)}
\end{eqnarray}
at $0\le t\le\tau$ and

\begin{equation}
\label{eq:a_l^TCMT_t>tau}
a_\ell^{\rm (TCMT)}(t)=\frac{i \gamma \left(e^{-\gamma\tau - i(\omega-\omega_0)\tau}-1\right) e^{ i[(\omega - \omega_0)t + \phi]+\gamma
t}}{\omega -\omega_0 -i\gamma},
\end{equation}
at $t>\tau$ (remember that $\gamma<0$).
Note, that $a_\ell^{\rm (TCMT)}(t)$ given by Eqs.~\eqref{eq:a_l^TCMT_t<tau}, \eqref{eq:a_l^TCMT_t>tau} are indeed slowly-varying relative
to $\exp(-i\omega t)$ since in the vicinity of a high-$Q$ resonance \mbox{$|\omega-\omega_0| \ll \omega$ and \mbox{$|\gamma| \ll \omega$}.}

Another point to be stressed is that $a_\ell^{\rm (TCMT)}(t)$ given by Eqs.~\eqref{eq:a_l^TCMT_t<tau},~\eqref{eq:a_l^TCMT_t>tau} is not
continuous at $t=0$ and $t=\tau$: At $t=0$ it has a jump from zero at $t=-0$ to $(\exp[i\phi]-1)/2$ at $t=+0$. At $t=\tau$ the jump has the
same value but the opposite sign. These discontinuities are a direct consequence of Eq.~\eqref{eq:TCMT_s}, which implies that the
background scattering follows the variations of $s^+(t)$ instantaneously, without any delay.

\subsection{Harmonic oscillators}
Another model is based on the well-known fact that any linear
oscillatory dynamic may be approximated by that of a system of driven coupled harmonic oscillators. We just have to apply it to the problem in question. The general solution of the
equations for driven coupled HO has the form (see, e.g.~\cite{LL_mechanics})
\begin{eqnarray}
z_{k}(t) = z_{ks}(t)+
\sum_{n}{C_n \Delta_{nk} e^{-i\hat{\omega}_n t}} \label{eq:HO_general_sol},
\end{eqnarray}
where $z_{k}(t)$ is a complex coordinate of the $k$-th oscillator; $\Delta_{nk} e^{-i\hat{\omega}_n t}$ and $\hat{\omega}_n$ stand for the
corresponding eigenvector and complex eigenfrequency, respectively; $z_{ks}(t)$ is a particular solution, for a given drive, and $C_n$ are
the constants of integration defined by the initial conditions. An important point is that if the eigenvectors are selected as new
variables (normal modes), the corresponding system of equations is diagonalized, i.e., each term in the sum in
Eq.~\eqref{eq:HO_general_sol} evolves \emph{independently\/} of the others and its dynamic is described by that of a \emph{single}
oscillator~\cite{LL_mechanics}.

The main idea of the adaptation of Eq.~\eqref{eq:HO_general_sol} for a drive with a carrier frequency $\omega$ is that only the dynamics of
the resonant eigenmodes with the frequency mismatch $|\omega - {\rm Re}\,\hat{\omega}_n|$ of the order of $|{\rm Im}\,\hat{\omega}_n|$ or
smaller than that are modeled. All other off-resonant eigenmodes are supposed to follow the drive adiabatically, obeying the quasi-steady approximation. This approach is reasonable, provided the contribution of the resonant modes to the overall dynamic is overwhelming.
Fortunately, this is the case in most resonant phenomena in subwavelength optics and related problems.

Next, according to what just has been {mentioned,} for the problem in question the eigenmode dynamic is described by the equation:
\begin{equation}\label{eq:driven_HO}
\ddot{f} - 2\gamma\dot{f} +\omega_{0}^2 f = A_0[\theta(t)-\theta(t-\tau)]\exp[-i\omega t];\;\; (\gamma<0),
\end{equation}
supplemented with the initial conditions \mbox{$f(0)=\dot{f}(0)=0$,} where dot stands for d/d$t$ and $\theta(z)$ is the Heaviside step
function.

Eq.~\eqref{eq:driven_HO} has the following exact solution:
\begin{eqnarray}
f(t) &=& A_0 e^{-i\omega t} \label{eq:fon} \frac{e^{(i\omega+\gamma)t}\left[\omega_{0\gamma}
\cos(\omega_{0\gamma}t)-
\left(\gamma+i\omega\right)
\sin(\omega_{0\gamma}t)+\right]-\omega_{0\gamma}}
{\left(\omega^2-\omega_0^2-2i\omega\gamma\right)
\omega_{0\gamma}},
\end{eqnarray}
at $0\leq t \leq \tau$ and
\begin{eqnarray}
f(t) &=& \frac{e^{\gamma(t-\tau)}}{\omega_{0\gamma}}
\Bigg\{\left[\dot{f}(\tau)-\gamma f(\tau)\right]
\sin\omega_{0\gamma}(t-\tau)+ \omega_{0\gamma}f(\tau)\cos\omega_{0\gamma}(t-\tau)\Bigg\}, \label{eq:foff}
\end{eqnarray}
at $t>\tau$.
Here $\omega_{0\gamma} \equiv \sqrt{\omega_0^2-\gamma^2}$.

{In what follows, we employ Eqs.~\eqref{eq:fon}, \eqref{eq:foff} to model the dynamics of the scattering coefficients. Naturally, for
different coefficients the values of $A_0,\;\omega_0$ and $\gamma$ are also different and will be defined below.} The key equation now is
Eq.~\eqref{eq:al-dl}, where the steady-state $a_\ell^{\rm (PEC)}$ and $d_\ell$ are replaced by $a_\ell^{\rm (PEC)}(t)$ and $d_\ell(t)$; and both are
regarded as independent eigenmodes.

Let us stress the dramatic difference in the steady-state profiles $|d_\ell(x)|$ and $|a_\ell^{\rm (PEC)}(x)|$. If for the former the
characteristic scale is of the order of $1/(nx) \ll 1$, for the latter it is just $1/x = O(1)$, see Eq.~\eqref{eq:al-dl} and
Ref.~\cite{Tribelsky_Mirosh_2016}. As a result, though in the vicinity of the maxima, the profiles $|d_\ell(x)|$ may be well-approximated
by Lorentzians, this is not the case for $|a_\ell^{\rm (PEC)}(x)|$. Therefore, the actual dynamic of each PEC-mode may be quite far from
that of a harmonic oscillator. Fortunately, since the characteristic scale in the time-domain is inverse of that in the
\mbox{$\omega$-domain}, the specified hierarchy of the scales means that the transients of the PEC-modes to the quasi-steady-state
scattering is fast. In contrast, the ones for the resonant partitions are relatively slow. Accordingly, the approximation of the latter
requires maximal accuracy. Regarding the possible errors in the approximation of the dynamics of the PEC-modes, they are not crucial to the
approach since due to the fastness of the PEC-modes, they affect just the very initial stage of the transient (a few periods $2\pi/\omega$, see below).

To employ Eq.~\eqref{eq:driven_HO} for modeling the dynamics of the eigenmodes, we have to fix the values of the following four parameters:
Re$[A_0]$, Im$[A_0],\;\omega_0$ \mbox{and $\gamma$.} \emph{The approximation details\/} are as follows: For the resonant \mbox{\textit{d}-modes} the values of $\omega_0$ and $\gamma$ are given
by the corresponding poles of $d_\ell(\omega)$ in the same manner as that in TCMT. Then, we require that at the drive frequency $\omega$,
the complex amplitude of the oscillator coincides with that for $d_\ell(\omega)$ of the exact solution.

{The profile of a PEC-mode is far from a narrow resonant line. Therefore, the poles in the complex plane might have nothing to do with the
dynamic of the mode. Then, $\omega_0$ and $\gamma$ become free parameters, and we need two more conditions to complete the problem. For them, we select (i) the equality of the frequencies maximizing the profile $|a^{\rm(PEC)}(\omega)|$ and the corresponding profile of the oscillator and (ii) the equality of the maxima themselves. This procedure fixes all four parameters of the HO-model unambiguously. For more details, see Appendix.}

\begin{figure}
\centering
\includegraphics[width=.9\columnwidth]{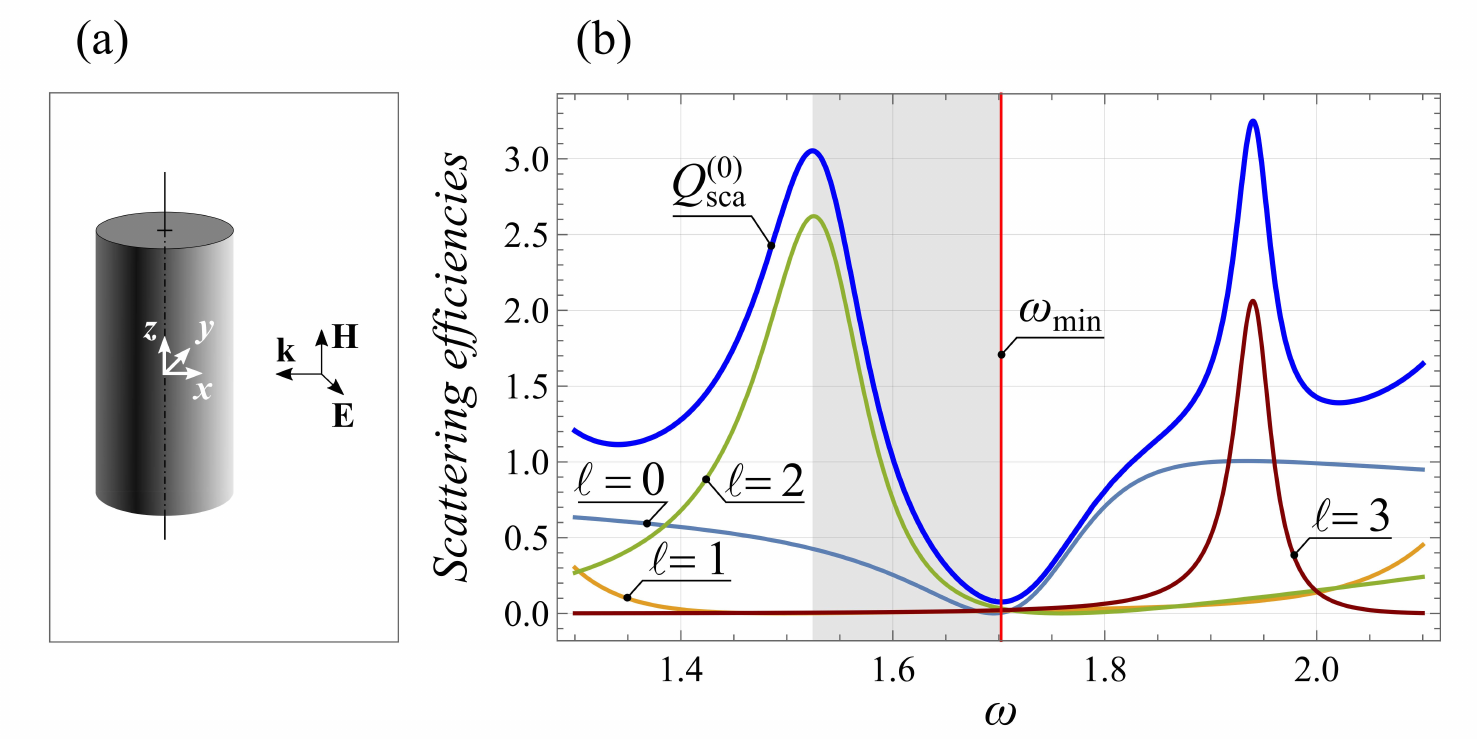}
\caption{{(a) The mutual orientation of the cylinder, coordinate frame, and vectors $\mathbf{k}$, $\mathbf{E}$, $\mathbf{H}$ of the incident linearly polarized plane wave. (b) Spectrum of the conventional total scattering efficiency, $Q_{\rm sca}^{(0)}$ and four first partial contribution to it: $Q_{{\rm sca}\,(0)}^{(0)}$ and $Q_{{\rm sca}\,(\ell)}^{(0)} + Q_{{\rm sca}\,(-\ell)}^{(0)}$ at $\ell \neq 0$, see  Eqs.\eqref{eq:Qsca}--\eqref{eq:Q_through_a}. The values of $\ell$ are indicated in the figure. Gray strip shows the range of dimensionless $\omega$ corresponding to Fig.~\ref{fig:precursor}. }
}\label{Efficiensies}
\end{figure}

\section{Results and Discussions}
\subsection{Parameters, variables and plots}

{ To illustrate the accuracy of the two models, their advantages and disadvantages, we perform computer simulations of the scattering based on the direct numerical integration of the complete set of Maxwell's equations with the help of Lumerical’s FDTD SOLUTIONS and compare their results with the dynamics described by the models. As it has been mentioned above, the mutual orientation of the cylinder and the incident wave corresponds to that shown in Fig.~\ref{Efficiensies}(a).}
%\del{comparison of their applications to the problem numerically studied in Ref.} %~\cite{Tribelsky_Mirosh:PRA_2019}.
It is convenient to transfer
to the dimensionless time: $t_{\rm new}=t_{\rm old}c/R$. Then, \mbox{$\omega_{\rm new}=\omega_{\rm old}R/c \equiv x$}. Since below only the
dimensionless quantities are in use, the subscript "new" will be dropped.

{The refractive index is selected purely real and equal to \mbox{$m=n=3.125$}, which corresponds to that of gallium phosphide ($GaP$) irradiated in a vacuum at the wavelength $\lambda^{\rm (GaP)}=966.4$ nm~\cite{refractiveindex.info}. For the size parameter the value \mbox{$x=1.702$} is selected. At the given wavelength, it implies that the radius of the cylinder equals $R=261.8$ nm. }
The choice is done since this pair of $m$ and $x$ corresponds to a local
minimum of $Q_{\rm sca}^{(0)}$ associated with the destructive Fano resonances at $\ell=0,\;\pm 2$, see Fig.~\ref{Efficiensies}(b), both of which are situated at the close
vicinity of $x=x_{\rm min}$ \mbox{($x \approx 1.695$} for \mbox{$\ell = 0$} and \mbox{$x \approx 1.759$} for \mbox{$\ell = \pm 2$)}. {Accordingly, the specified above values of $m,\;x,\;\lambda$, and $\omega = x$ the subscript "min" is assigned. }

For all other multipoles, $x=x_{\rm min}$ corresponds to the off-resonant regions. Thus, only the dynamics of the modes with $\ell=0,\;\pm 2$
should be approximated. Moreover, since the scattering coefficients differing only by the sign of $\ell$ are identical, the modes with
$\ell=\pm 2$ may be regarded as a single one.

{The pulse duration, $\tau$ equals 191.28. This is much larger than any problem's characteristic relaxation time at the given values of the other parameters. Then, it makes it possible to study both: the transient to the stationary scattering (behind the leading edge of the incident pulse) and its decay (behind the trailing edge).}

The poles of the scattering coefficients adjacent to $\omega=\omega_{\rm min}$ are
\begin{equation}\label{eq:poles}
\hat{\omega}_0^{\ell=0}\approx 1.741-0.097i;\;\;
\hat\omega_0^{\ell=2}\approx 1.535-0.0614i;
\end{equation}
{Then, the described above procedure gives rise to the following values of the models' parameters (see Appendix for details):}

\begin{figure}
\centering
\includegraphics[width=.9\columnwidth]{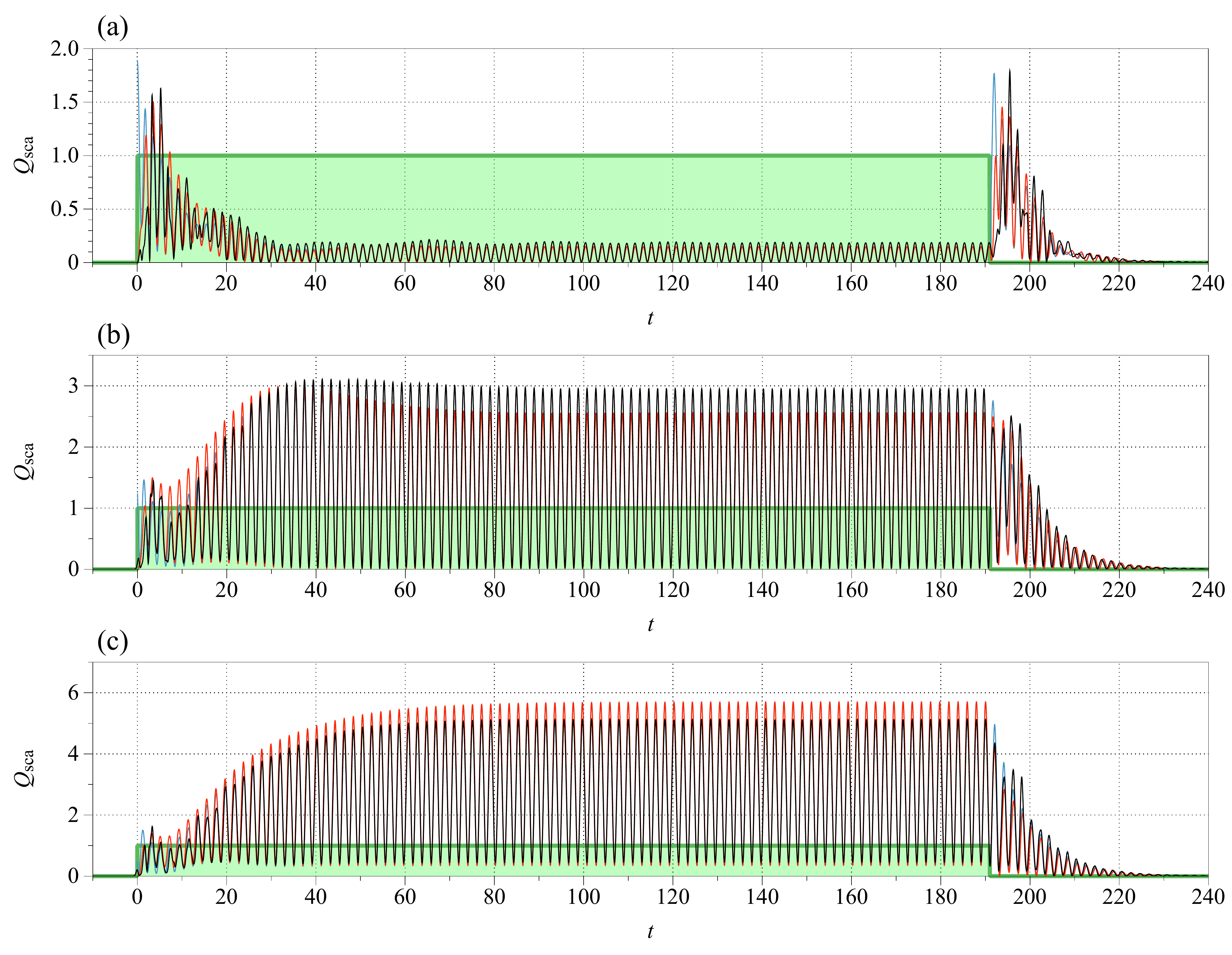}%[h]
\caption{The full scattering dynamics of $GaP$ cylinder of radius $R=261.8$ nm {and its approximation by the two models} in various regimes: (a) at the destructive Fano resonance condition: $\omega =\omega_{\rm min}= 1.702$ ($\lambda^{\rm (GaP)}=966.4$ nm); (b) at the intermediate wavelength: $\omega =\omega_{\rm mid}= 1.589$ ($\lambda^{\rm (GaP)}=1035$ nm); and (c) at the {local} resonant maximum {of the} scattering: $\omega =\omega_{\rm max}= 1.525$ ($\lambda^{\rm (GaP)}=1079$ nm). In each panel the direct numerical results are shown by black curve, TCMT by blue curve, and HO by red curve. The variation in the steady-state response is clearly seen. Both models give very good agreement with the numerical simulation. {Normalized incident pulse is shown as a filled green rectangle.}}\label{fig:all_dynamics}
\end{figure}

\begin{figure}%[h]
\centering
\includegraphics[width=.7\columnwidth]{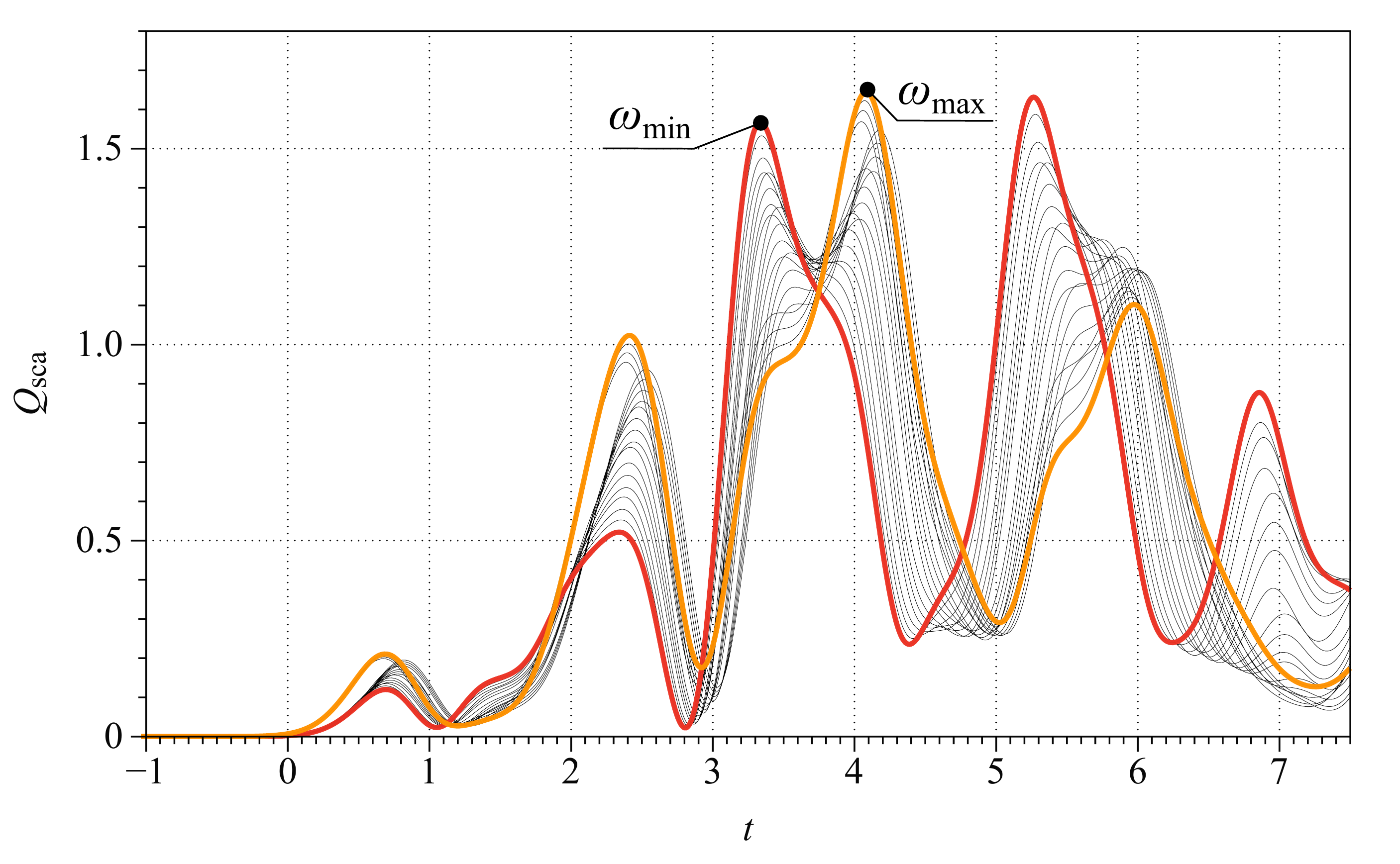}
\caption{{Set of plots exhibiting the dynamics of the precursor %\del{during the initial excitation just behind the leading edge of the incident pulse in}
at different carrier frequencies of the incident light} obtained by direct numerical integration of Maxwell's equations. {The dimensionless frequency varies from $\omega = \omega_{\rm max} = 1.525$ to $\omega =\omega_{\rm min}= 1.702$ with step \mbox{$\Delta \omega = 0.00885$.} At the selected values of the problem parameters for $GaP$ this corresponds to the variation of the wavelength from 966.4~nm to 1079~nm, cf. Fig.~\ref{fig:all_dynamics}.}
}\label{fig:precursor}
\end{figure}

\noindent\emph{For TCMT}:
\begin{equation}\label{eq: param_phi}
\phi^{\ell=0}(\omega_{\rm min})\approx -2.263,\;\; \phi^{\ell=2}(\omega_{\rm min})\approx 0.471,
\end{equation}

\noindent\emph{For HO}:

$d$-modes: $\hat{\omega}_\ell$ are given by Eq.~\eqref{eq:poles}. Regarding $A_0$,
\begin{equation}\label{eq: param_d}
A^{\ell=0}_{0 (d)}\! \approx\! -0.263 \!+\!0.554i;\;\; A^{\ell=2}_{0 (d)}\! \approx\! 0.555 \!+\! 0.142i.
\end{equation}

PEC-modes:
\begin{eqnarray}
& &\!\!\!\!\!\!\!\!\!\! \hat{\omega}_{0{\rm (PEC)}}^{\ell=0}\! \approx\! 2.495\! -\! 0.837i;\;\;
\hat\omega_{0{\rm (PEC)}}^{\ell=2}\!\approx\! 2.137\! -\! 0.532i; \label{eq:PEC_OMEGA}\\
& &\!\!\!\!\!\!\!\!\!\! A^{\ell=0}_{0{\rm (PEC)}}\! \approx\! 3.811\! - \!0.979i;\;\; A^{\ell=2}_{0{\rm (PEC)}} \! \approx\! -0.392\! -\!
0.793i.\label{eq:PEC_An}
\end{eqnarray}
{Note that, despite the error in the approximation of $a^{\rm (PEC)}$, the approximation of $a_\ell(\omega)$ in the vicinity of the Fano
resonances is quite accurate, cf. Figs.~\ref{fig:aPEC}, \ref{fig:a}.}

{The quantitative comparison of the two models with the numerics %\del{of Ref.~[1]}
at the dimensionless $\omega=\omega_{\rm min}$, shown in
%Fig.~\ref{Efficiensies}(b,c)
Fig.~\ref{fig:all_dynamics}, exhibits their high accuracy. The origin of the $t$-axis in Fig.~\ref{fig:all_dynamics} (and below, in Figs.~\ref{fig:precursor}, \ref{fig:directional_scattering}) is shifted so that $t=0$ corresponds to the moment
when the scattered radiation for the first time is detected by the measuring monitors, situated in the far wave zone.} %\del{The comparison

{ To demonstrate that the developed models are applicable to describe with high accuracy the dynamic scattering not only in the vicinity of the minimum of the Fano line but within the entire Fano resonance line, as a whole, we also show in Fig.~\ref{fig:all_dynamics},
their comparison with the numerics at the point of the local maximum of the steady-state scattering ($\omega=\omega_{\rm max}=1.525;\;\lambda^{\rm (GaP)}=1079$~nm) and in the middle of the line ($\omega=\omega_{\rm mid}=1.589;\;\lambda^{\rm (GaP)}=1035$~nm), see Fig.~\ref{Efficiensies}(b)}; {the meaning of the subscripts {\it min\/} and {\it max\/} is connected with the scattering intensity, therefore, please do not be confuse by the fact that numerical value of $\omega_{\rm max}$ is smaller than that for $\omega_{\rm min}$.}

{The number of dominant modes in these cases remains the same as that at $\omega=\omega_{\rm min}$. Therefore, the poles and resonant conditions of each mode do not change too. The fitting procedure remains identical to that described above, though, of course, the matching conditions should be implemented for the corresponding new values of the driver frequency.}

{Now, let us discuss the most interesting, counterintuitive effect in the scattering dynamics, namely the sharp intensive spikes just behind the leading ({\it precursor}) and trailing ({\it postcursor)}  edges of the incident pulse, observed at $\omega=\omega_{\rm min}$ both in the total and directional scattering, see Figs.~\ref{fig:all_dynamics}(a), and \ref{fig:directional_scattering}.}
{Note that while the pre- and postcursor look similar, their nature is completely different. The precursor is related to the fast excitation of the PEC-modes with the low $Q$-factors. It lasts for the period while the relatively slow resonant modes are not "swung" yet. Since the line of any PEC-mode is broad, the variation of the carrier frequency of the incident pulse in the range $1.525 \leq \omega \leq 1.702$ affects the value of the PEC-line profile very little, see Fig.~\ref{fig:aPEC}. Accordingly, it weakly affects the precursor dynamics; see Fig.~\ref{fig:precursor}, where we show a set of the precursor lines in the discussed range of $\omega$ variations with the step $\Delta\omega = 0.00885$. For all curves shown in Fig.~\ref{fig:precursor} the
maximal value of the precursor is about 1.6, and its shape does not change much.}

\begin{figure}[h]
\includegraphics[width=.9\columnwidth]{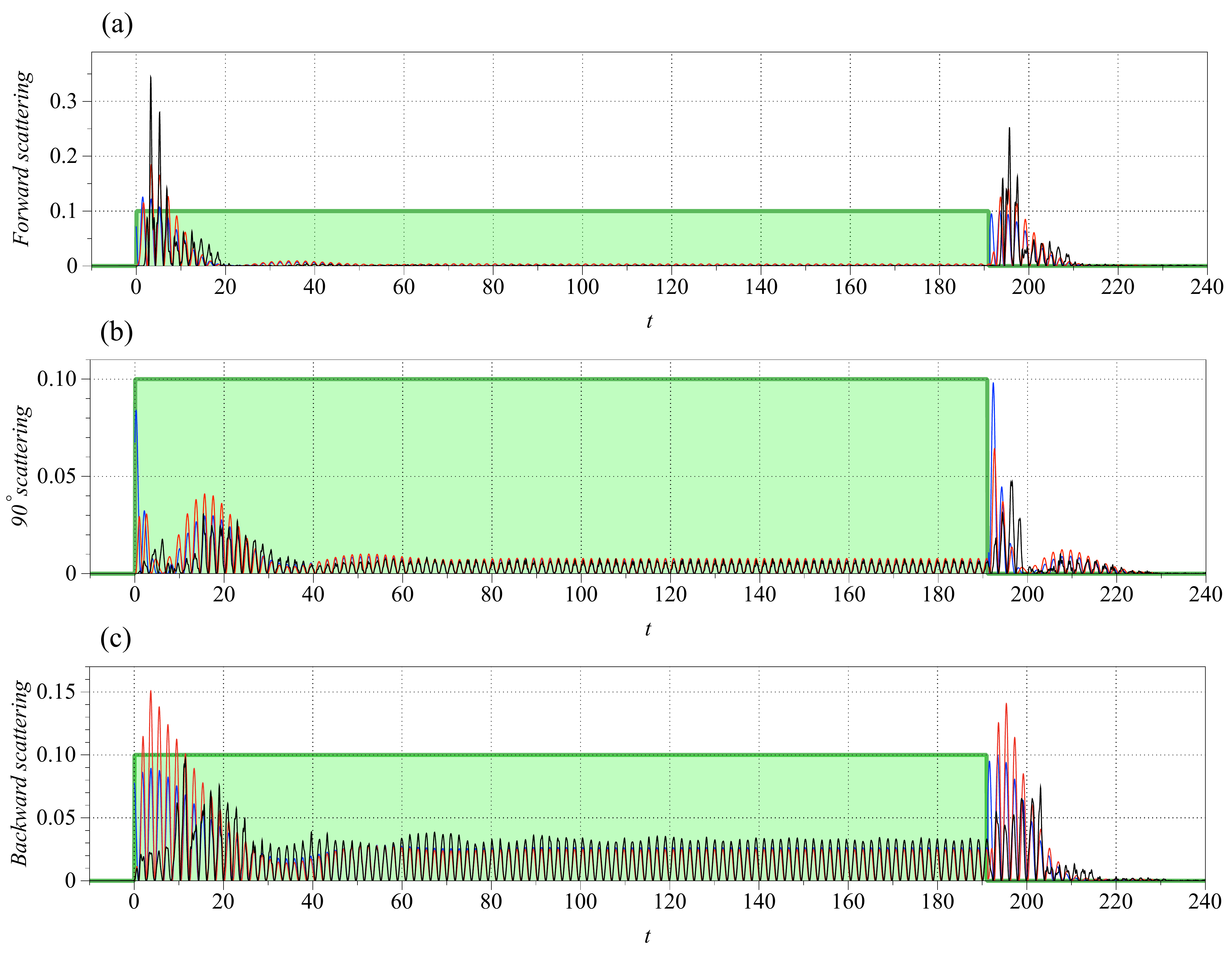}
\centering
\caption{{Directional scattering dynamics at the Fano destructive interference condition for $GaP$ at \mbox{$\omega =\omega_{\rm min}= 1.702$} ($\lambda^{\rm (GaP)}=966.4$ nm): (a) forward scattering, (b) $90^\circ$ scattering, and (c) backward scattering along the incident polarization. The direct numerical results are presented in each panel as a black curve, TCMT as a blue curve, and HO as a red curve. The incident pulse is shown as a filled green rectangle. The intensity of the scattering is normalized over the intensity of the incident light. To make it comparable with the intensity of the scattering, the pulse's intensity is reduced to 0.1 of its actual value. Note that the positions of the precursor for 90$^\circ$ scattering is shifted with respect to those for forward and backward scattering.}}\label{fig:directional_scattering}
\end{figure}

{In contrast, the postcursor is explained by the violation of the balance between the resonant and PEC-modes caused by the rapid decay of the latter, when the pumping with the incident pulse is over, and electromagnetic energy of the dark, non-radiating in the steady-state scattering modes, begins to leak out. Thus, the maximal value of the postcursor is bounded by the amplitude of the resonant modes just behind the trailing edge of the incident pulse. Due to the narrowness of the resonant lines, the same variation of $\omega$ ($1.525 \leq \omega \leq 1.702$) affects the steady-state amplitude of the resonant modes substantially, see Fig.~\ref{fig:d}. Accordingly, at the point of the destructive interference at $\omega = 1.702$, where the steady-state amplitudes of the resonant and background partitions are almost equal, the maximal values of the pre- and postcursor are almost equal too. Regarding the duration of the pre- and postcursor, both are determined by the relaxation times of the resonant modes (i.e., by $1/\gamma$), which, therefore, are the same for the pre- and postcursor. As a result, the shapes of the pre- and postcursor in Fig.~\ref{fig:all_dynamics}(a) are very similar to each other.}

{However, a departure from $\omega = 1.702$ toward $\omega = 1.525$ increases the steady-state amplitude of the resonant partition, while the one for the background remains practically unchanged. This gives rise to a decrease in the ratio of the maximal value of the postcursor to the amplitude of the steady-state scattering. Eventually, at the local maximum of the scattering, the "basement" of the stationary scattering eats out the postcursor almost entirely, cf. Fig.~\ref{fig:all_dynamics}(a) and \ref{fig:all_dynamics}(c).}

{It is important to note that both developed models make it possible to describe the dynamic of each multipole independently. Therefore, in addition to the reproduction of the total dynamic scattering, the models can be equally applied to describe the directional dynamic scattering too. The latter may be more practical from the viewpoint of getting experimental evidence of the discussed effects.}

{Qualitatively, the directional scattering occurs similar to the overall one. In particular, in the vicinity of $\omega=\omega_{\rm min}=1.702$ ($\lambda^{\rm(GaP)}_{\rm min} = 966.4$~nm) the pre- and postcursor spikes are clearly exhibited in the directional scattering at an arbitrary scattering angle. As an example, in Fig.~\ref{fig:directional_scattering} we show the comparison of forward, backward, and  $90^\circ$ scattering at  $\omega=\omega_{\rm min}$. However, the difference in the relaxation times for various multipoles and their resonant/background partitions gives rise to dynamic alteration of the overall scattering radiation pattern. As a result, the momentary dominant scattering lobe changes from forward to backward and even transverse directions, see Fig.~\ref{fig:directional_scattering} .} {All these intriguing dynamic effects might lead to a plethora of new and exciting applications.}

\subsection{Root mean square errors}

{To get a quantitative criterion of the accuracy of the models in a given window \mbox{$t_1 \leq t \leq t_2$} we do the following: The time
in the window is sampled with a step $\delta t$. Then, the root mean square error (RMSE) is defined as follows
\begin{eqnarray}
RMSE^{\rm (mod)}(N_1,N_2) = \frac{1}{\sqrt{N_2-N_1}}\sum\limits_{n=N_1}^{N_2}\sqrt{\left\{Q_{\rm sca}^{\rm
(num)}(n\delta t)-Q_{\rm sca}^{\rm (mod)}(n\delta t)\right\}^2}
\end{eqnarray}
is calculated} {(do not mix dummy $n$ here with the real part of the complex refractive index $m$).} {Here $Q_{\rm sca}^{\rm (num,\;mod)}(t)$ designates the corresponding
dynamic scattering for the numerics and each model, respectively; while integers $N_1$ and $N_2$ define start and end times: $t_1=N_1\delta t$ and $t_2=N_2\delta t$. At \mbox{$\delta t \rightarrow 0$} the RMSE converges to a certain constant $RMSE(t_1,t_2)$,
regarded as a measure of the model accuracy in the given window.}
\begin{table}[!ht]
\caption{Calculated total dynamic scattering cross sections RMSE for two models  for precursor and postcursor at three excitation frequencies, $\omega_{\rm min}=1.702$, $\omega_{\rm mid}=1.589$, and $\omega_{\rm max}=1.525$.}
\label{tab:RMSE_total}
\end{table}
\begin{center}
\begin{minipage}[t]{0.7\textwidth}
\begin{ruledtabular}
\begin{tabular}{c cc||cc}
\multicolumn{5}{c}{RMSE for total scattering cross section} \\ %\midrule
\hline
 & HO & TCMT & HO & TCMT \\
\hline
$\omega=\omega_{\rm min}$ & 0.144 & 0.181 & 0.114 & 0.166	\\
$\omega=\omega_{\rm mid}$ & 0.282 & 0.211 & 0.192 & 0.273	\\
$\omega=\omega_{\rm max}$ & 0.295 & 0.252 & 0.298 & 0.212	\\
\hline%\midrule
& \multicolumn{2}{c}{Precursor} & \multicolumn{2}{c}{Postcursor} \\
%\midrule
\end{tabular}
\end{ruledtabular}
%\end{adjustbox}
%\end{table}
\end{minipage}
\end{center}
\begin{table}[!ht]
\caption{Calculated directional dynamic scattering RMSE for two models for precursor and postcursor including forward, backward, and $90^\circ$ directions at $\omega=\omega_{\rm min}=1.702$.}
\label{tab:RMSE_directional}
\end{table}
\begin{center}
\begin{minipage}[t]{0.7\textwidth}
\begin{ruledtabular}
\begin{tabular}{c cc||cc}
\multicolumn{5}{c}{RMSE for directional 90$^\circ$ scattering} \\
\hline %\midrule
 & HO & TCMT & HO & TCMT \\
\hline %\midrule
Forward & 0.0295 & 0.0261 & 0.0265 & 0.0285	\\
Backward & 0.0291 & 0.0236 & 0.0188 & 0.0189	\\
90$^\circ$ & 0.0066 & 0.0130 & 0.0063 & 0.0144 \\
\hline
&\multicolumn{2}{c}{Precursor} & \multicolumn{2}{c}{Postcursor} \\
\end{tabular}
\end{ruledtabular}
\end{minipage}
\end{center}

{For the two windows with essentially non-steady dynamics, adjacent to the
leading and trailing edges of the incident pulse ($t_1=0,\;t_2=38$ and $t_1=\tau,\;t_2=\tau+38$)\footnote{The selected width of the
window $\Delta t = 38$ satisfies the condition $\exp(-|\gamma| \Delta t) \approx 0.1$, where $|\gamma|$ is the smallest decrement
corresponding to $\ell=2$, see Eq.~\eqref{eq:poles}.}, the calculated errors are summarized in Table~\ref{tab:RMSE_total} for the total dynamic scattering and in Table~\ref{tab:RMSE_directional} for the directional one. The comparison of the results for the precursor $RMSE^{\rm (TCMT)}=0.181$, $RMSE^{\rm
(HO)}=0.144$, and the postcursor \mbox{$RMSE^{\rm (TCMT)}=0.166$,} \mbox{$RMSE^{\rm (HO)}=0.114$,} in the vicinity of the destructive Fano interference conditions ($\omega=\omega_{\rm min}=1.702$) illustrates that the HO model is more accurate. This is expected since the dynamics of each mode is modeled by the second-order equation instead of the first one (like in the TCMT case), allowing to capture the transient dynamics in greater details. }

\section{Conclusions}
Thus, though both models demonstrate high accuracy, the one of the HO is a little better than that of TCMT. In addition, while the
TCMT-model just \emph{describes\/} the complicated dynamics of the resonant light scattering,  the HO-model also elucidates the physical
grounds for the sharp intensive spikes in the scattering observed behind the leading and trailing edges of the incident
pulse and explained by the violation of the balance between the resonant and background excitations. The
violation, in turn, is coursed by the difference in the corresponding relaxation times. Then, the uncompensated part of the interfering
modes is exhibited as a spike.}

Note that if the "basement" of the scattered pulse is cut off at the level above the intensity of the
steady-state scattering, the duration of the remaining parts of the spikes occurs very small. {This procedure may be used as a new method of radiation at nanoscale of sharply-directed pulses with a duration of a few periods of the field oscillations.
}

{We have also shown that both models can be successfully applied to describe the total and directional dynamic scattering at an arbitrary wavelength. On top of that, importantly, the models also reveal the underlying physical phenomena associated with the fast and slow modes excitation, whose interference leads to new effects, including the formation of the sharp spikes.}

Regarding possible extensions of the models, while the governing equations of TCMT are simpler than those of HO, the accuracy of the
HO-model is higher relative to the one of TCMT. Besides, the implementation of the HO model is more straightforward and does not require a
sophisticated procedure to connect the model's parameters with those of the initial underlying problem.

If just a single resonant excitation is a concern, the HO-model may be used based even on fitting an experimentally obtained spectrum of
the output signal modules. On the other hand, when the interference of several excitations is essential, its application implies knowledge of the phases, obtained either from an analytical solution or measured experimentally. While the latter is challenging in the optical range, it is a routine procedure at radio frequencies; see,
e.g.,~\cite{Tribelsky2015,PRB:DirectionaFano2016} and references therein. Thus, both models complement each other and maybe very useful in
descriptions of a wide variety of resonant phenomena.

\begin{acknowledgements}
  The authors are very grateful to Boris Y. Rubinstein for his valuable help in symbolic computer calculations.
\end{acknowledgements}

%\begin{funding}
M.I.T. acknowledges the financial support of the Russian Foundation for Basic Research (Projects No. 20-02-00086) for
the analytical study, the Moscow Engineering Physics Institute Academic Excellence Project (agreement with the Ministry of Education and
Science of the Russian Federation of 27 August 2013, Project No. 02.a03.21.0005) for the modeling of the resonant light scattering, as well
as the contribution of the Russian Science Foundation for the computer simulation (Project No. 21-12-00151) and the provision of user
facilities (Project No. 19-72-30012).
%\end{funding}
%\bibliographystyle{...}
%\bibliography{...}

% Add the appendix
%\appendix
%set the numbering of sections, figures, and equations to letters
\renewcommand\thefigure{\thesection.\arabic{figure}}
\renewcommand{\theequation}{\thesection.\arabic{equation}}
% reset the numbering of figures
\setcounter{figure}{0}
\setcounter{equation}{0}

\section{Appendix}

\subsection{Scattering coefficients}

The scattering coefficient for the problem under consideration read as follows~\cite{Bohren::1998}:

\begin{eqnarray}
a_\ell &=& \frac{mJ_\ell(mx)J^{\prime}_\ell(x) - J_\ell(x)J_\ell'(mx)}{mJ_\ell(mx)H^{(1)\prime}_\ell(x) - H^{(1)}_\ell(x)J_\ell'(mx)}, \label{eq:a_ell} \\
d_\ell &=& \frac{{2i}/{(\pi x)}}{mJ_\ell(mx)H^{(1)\prime}_\ell(x) - H^{(1)}_\ell(x)J_\ell'(mx)}\label{eq:d_ell}\;,
\end{eqnarray}
%see Eqs.~\eqref{eq:a_ell}--\eqref{eq:d_ell}. {In Fig.~\ref{fig:spectrum} the total, Eq. (\ref{eq:Qsca}), and partial, Eq. (\ref{eq:Q_through_a}), scattering contributions are shown. The minimal total scattering, corresponding to the Fano destructive interference, takes place at $\lambda=537$nm.}
%
%\begin{figure}[h]
%\centering
%\includegraphics[width=.7\textwidth]{Fig_spectrum.pdf}
%\caption{Total {($Q_{\rm sca}$)} and partial {$(Q^{(\ell)}_{\rm sca})$} scattering spectra of $GaP$ cylinder with radius $R=130$nm.}\label{fig:spectrum}
%\end{figure}
\subsection{The approximation procedure}
In what follows $\omega_{\rm sim}$ designates the carrier frequency of the incident pulse corresponding to a given simulation, while all specific examples are given at $\omega_{\rm sim}=\omega_{\rm min}=1.702$.   The application procedure at other values of $\omega_{\rm sim}$ discussed in the main text is identical to that at $\omega_{\rm sim}=\omega_{\rm min}$.
\subsubsection{TCMT-model}

{To obtain the value of $\phi$, note that according to Ref.~\cite{Fan:TCM_2010} $a_{\ell}^{\rm(TCMT)}$ may be presented as
\begin{eqnarray}\label{eq:a_ss}
\begin{aligned} a_{\ell}^{\rm(TCMT)} %&=\frac{1}{2}(r-1) \\ &
=\frac{1}{2} \frac{{i}\left(\omega_{0}-\omega\right) ({e}^{{i} \phi}-1 )+\gamma\left(1+{e}^{{i} \phi}\right)}{{i}\left(\omega_{0}-\omega\right)-\gamma}. \end{aligned}
\end{eqnarray}
(the same expression may be obtained from Eq.~(\ref{eq:a_l^TCMT_t<tau}), if we formally consider its limit at $t \rightarrow \infty$). Then, after some algebra Eq.~\eqref{eq:a_ss} gives rise to the following expression for $|a_{\ell}^{\rm(TCMT)}|^2$:
\begin{equation}\label{eq:Abs[a_ss]^2}
|a_{\ell}^{\rm(TCMT)}|^2 %& =\frac{|1-e^{i\phi}|^2}{4}\left| \frac{i\frac{\omega_0-\omega}{\gamma}+\frac{1+e^{i\phi}}{1-e^{i\phi}}}{i\frac{\omega_0-\omega}{\gamma}+1}\right|^2\\&
=\sin^2\frac{\phi}{2}\left| \frac{\cot\frac{\phi}{2}+\frac{\omega-\omega_0}{\gamma}}{1+i\frac{\omega-\omega_0}{\gamma}}\right|^2 =\frac{1}{1+q^2}\frac{(q+\epsilon)^2}{1+\epsilon^2},
\end{equation}
where $\epsilon=-(\omega-\omega_0)/\gamma$ and $ q=-\cot(\phi/{2})$. We remind that $\gamma<0$. The last formula in Eq.~\eqref{eq:Abs[a_ss]^2} is the conventional Fano profile with the asymmetry parameter $q$~\cite{Fano:NC:1935,Fano:PR:1961,Miroshnichenko:2010ewa}.}

{Following the procedure described in the main text, we have to equalize $|a_{\ell}^{\rm(TCMT)}(\omega_{\rm sim})|^2$ to $|a_{\ell}(\omega_{\rm sim})|^2$, where $a_{\ell}(\omega_{\rm sim})$ is given by Eq.~\eqref{eq:a_ell} at $x=\omega_{\rm sim}$ (remember that in the selected dimensionless variables $x$ and $\omega$ numerically are equal to each other). This brings about a quadratic equation for $q$, whose solutions are
\begin{equation}\label{eq:q_pm}
q_{_{(\pm)}} = \frac{\epsilon_{\rm sim}\pm(1+\epsilon_{\rm sim}^2)|a_{\ell}(\omega_{\rm sim})|\sqrt{1-|a_{\ell}(\omega_{\rm sim})|^2}}{(1+\epsilon_{\rm sim}^2)|a_{\ell}(\omega_{\rm sim})|^2-1},
\end{equation}
where $\epsilon_{\rm sim}=-(\omega_{\rm sim}-\omega_0)/\gamma$. Note that for the problem in question $|a_\ell|$ is always smaller than unity, see, e.g.,~\cite{Tribelsky:2013ft}. Therefore, the roots in Eq.~\eqref{eq:q_pm} are always real.}

\begin{figure}
\centering
\includegraphics[width=\textwidth]{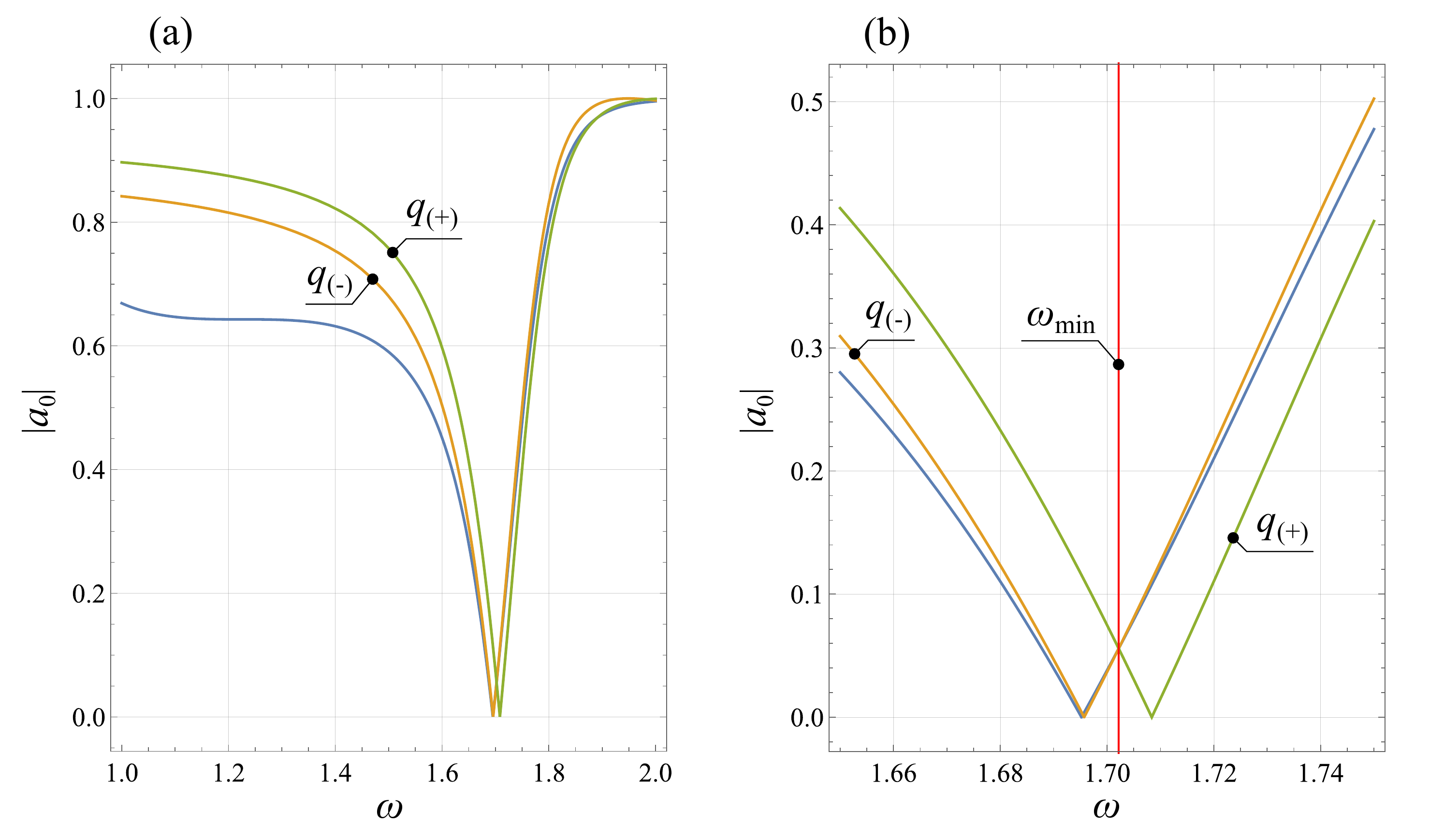}
\caption{The approximation of the profile $|a_0(\omega)|$ shown as a blue line (unmarked) by Eq.~\eqref{eq:Abs[a_ss]^2} at $q=q_{_{(-)}}$ and $q=q_{_{(+)}}$. The general view (a) and the vicinity of $\omega=\omega_{\rm min}$ (b). Though all plots have the same value at $\omega=\omega_{\rm min}$, the overall approximation is better at $q=q_{_{(-)}}$.}\label{fig:a0qpm}
\end{figure}

\begin{figure}
\centering
\includegraphics[width=.8\textwidth]{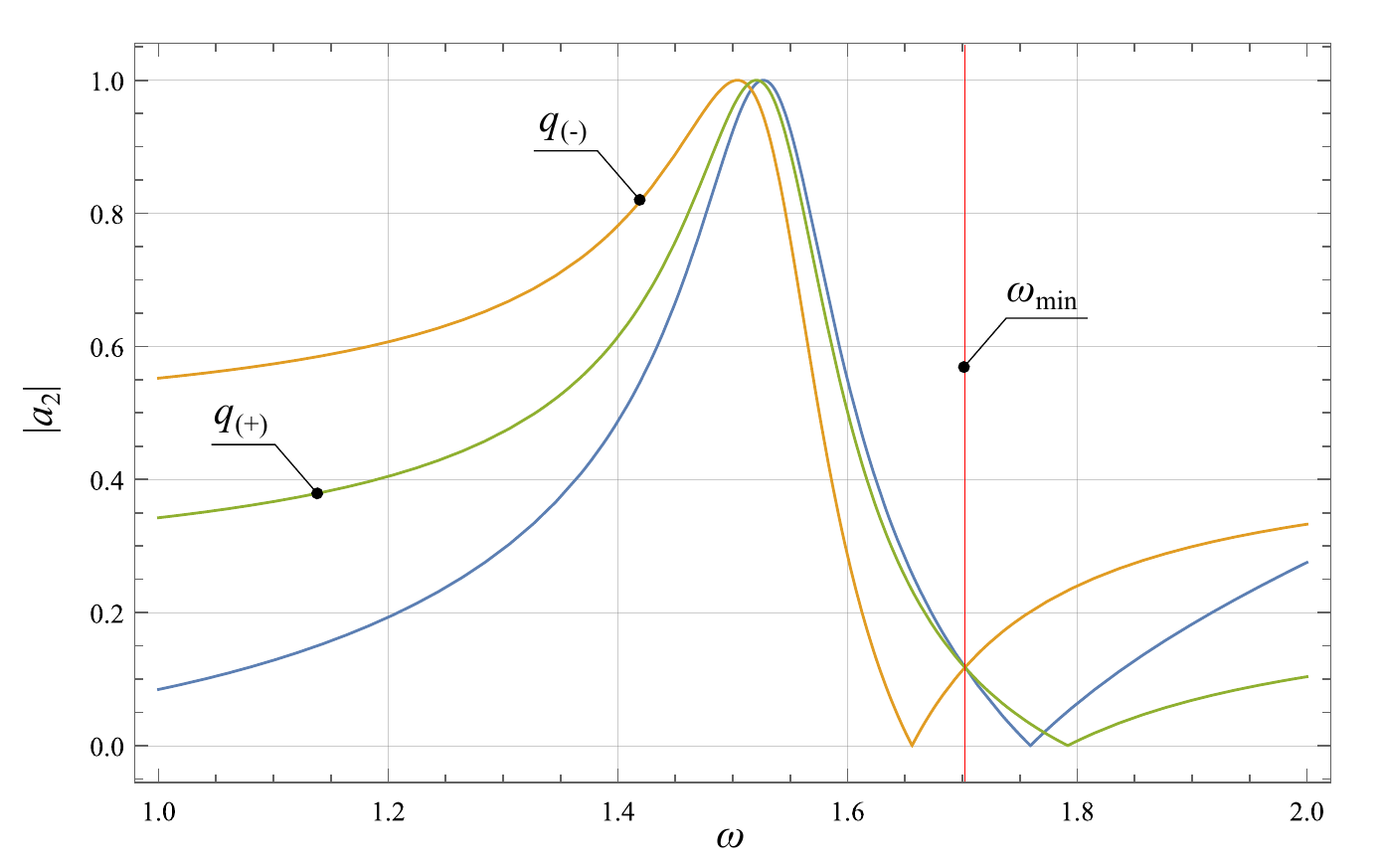}
\caption{The same as that in Fig.~\ref{fig:a0qpm}(a) for $\ell=2$. The better overall approximation is for $q=q_{_{(+)}}$}\label{fig:a2qpm}
\end{figure}

{Plots $|a_{\ell}(\omega)|$, as well as their approximations at $\omega_{\rm sim}=\omega_{\rm min}$, $q=q_{_{(+)}},\;q=q_{_{(-)}}$ and $\ell =0,\;2$ for the values of $\omega_0,\;\gamma$ given by Eq.~(\ref{eq:poles}), are shown in Figs.~\ref{fig:a0qpm}--\ref{fig:a2qpm}. Thus, for $\ell=0$, the better overall approximation is at $q=q_{_{(-)}}$. For the case in question its numerical value is 0.4700, which corresponds to $\phi \approx -2.263$. In contrast, for $\ell=2$, the better overall approximation is at $q=q_{_{(+)}} \approx -4.172$, corresponding to $\phi \approx 0.471$.}

\begin{figure}%[h]
\centering
\includegraphics[width=.9\textwidth]{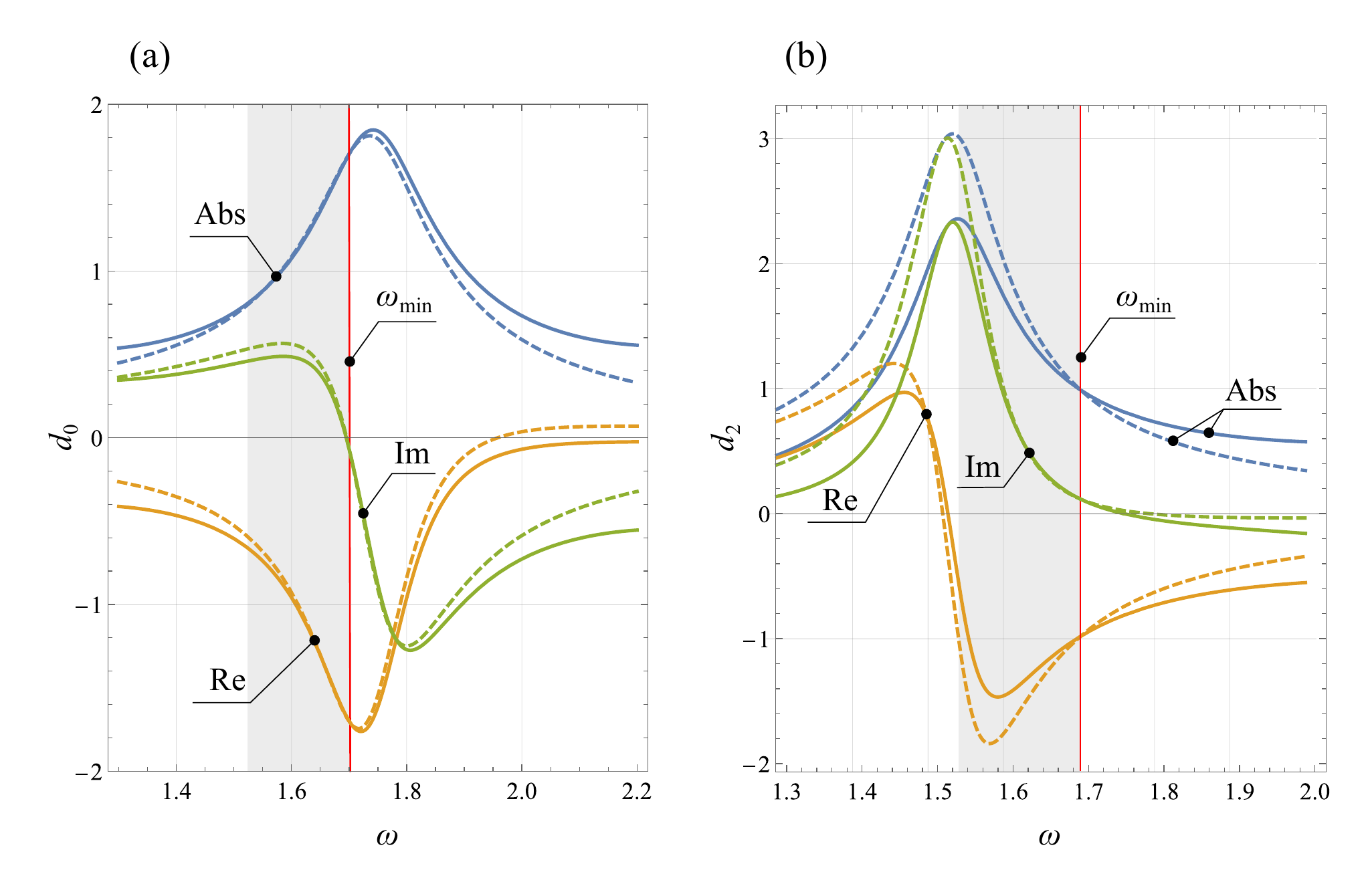}
\caption{$d$-modes. Exact steady-state scattering coefficients (solid lines) and their approximations in the HO-model (dashed lines) at $\omega_{\rm sim}=\omega_{\rm min}$. Abs, Re and Im designate the modula, real and imaginary parts of the coefficients, respectively. The larger the difference between $\omega_{\rm sim}$ and the resonant frequency maximizing $|d_\ell|$, the lager the approximation error, cf. panels (a) and (b). {Gray strip shows the range of $\omega$ corresponding to Fig.~\ref{fig:precursor}.}}\label{fig:d}
\end{figure}

\begin{figure}
\centering
\includegraphics[width=.9\textwidth]{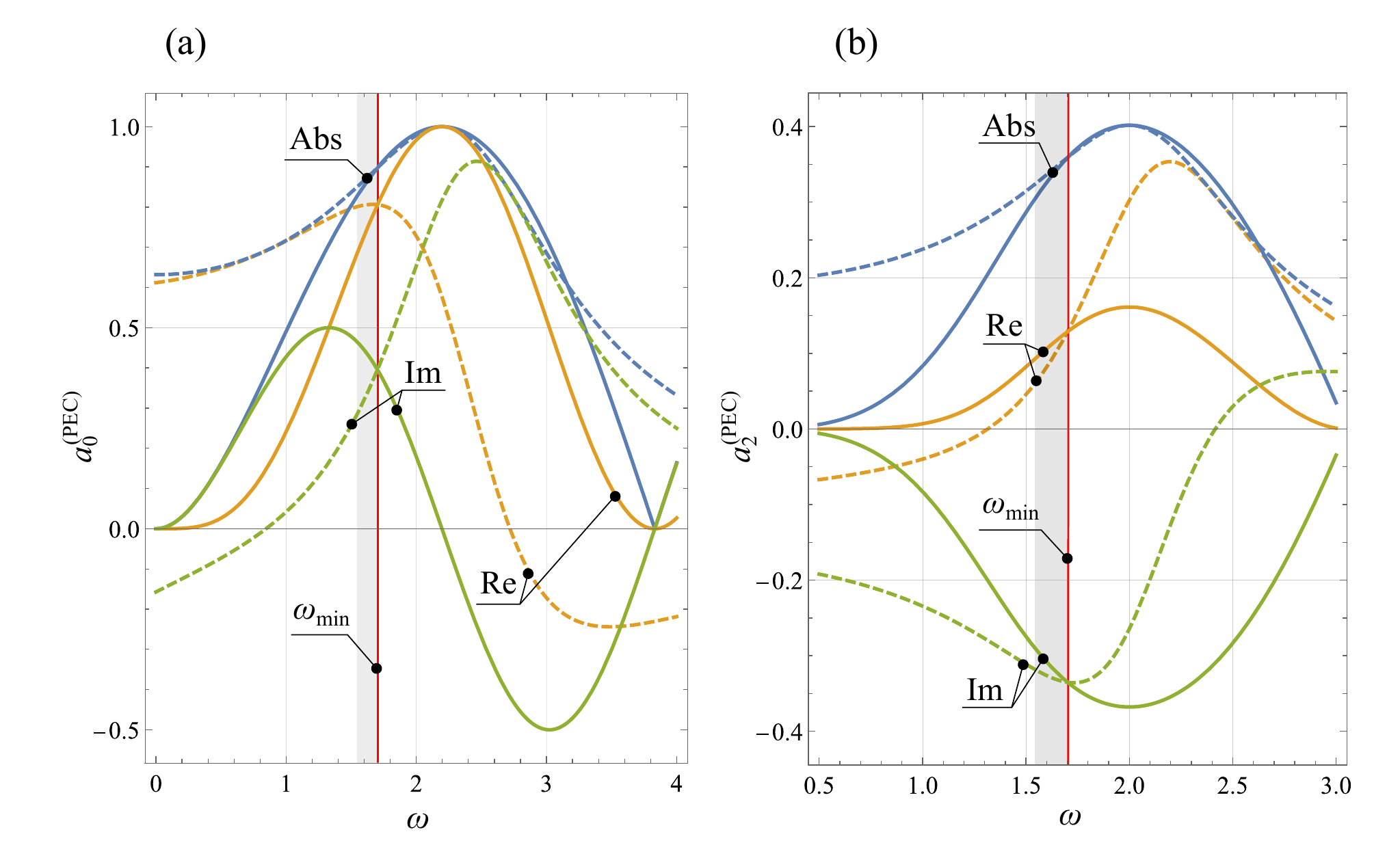}
\caption{The same as that in Fig.~\ref{fig:d} for PEC-modes. }\label{fig:aPEC}
\end{figure}

\begin{figure}
\centering
\includegraphics[width=.9\textwidth]{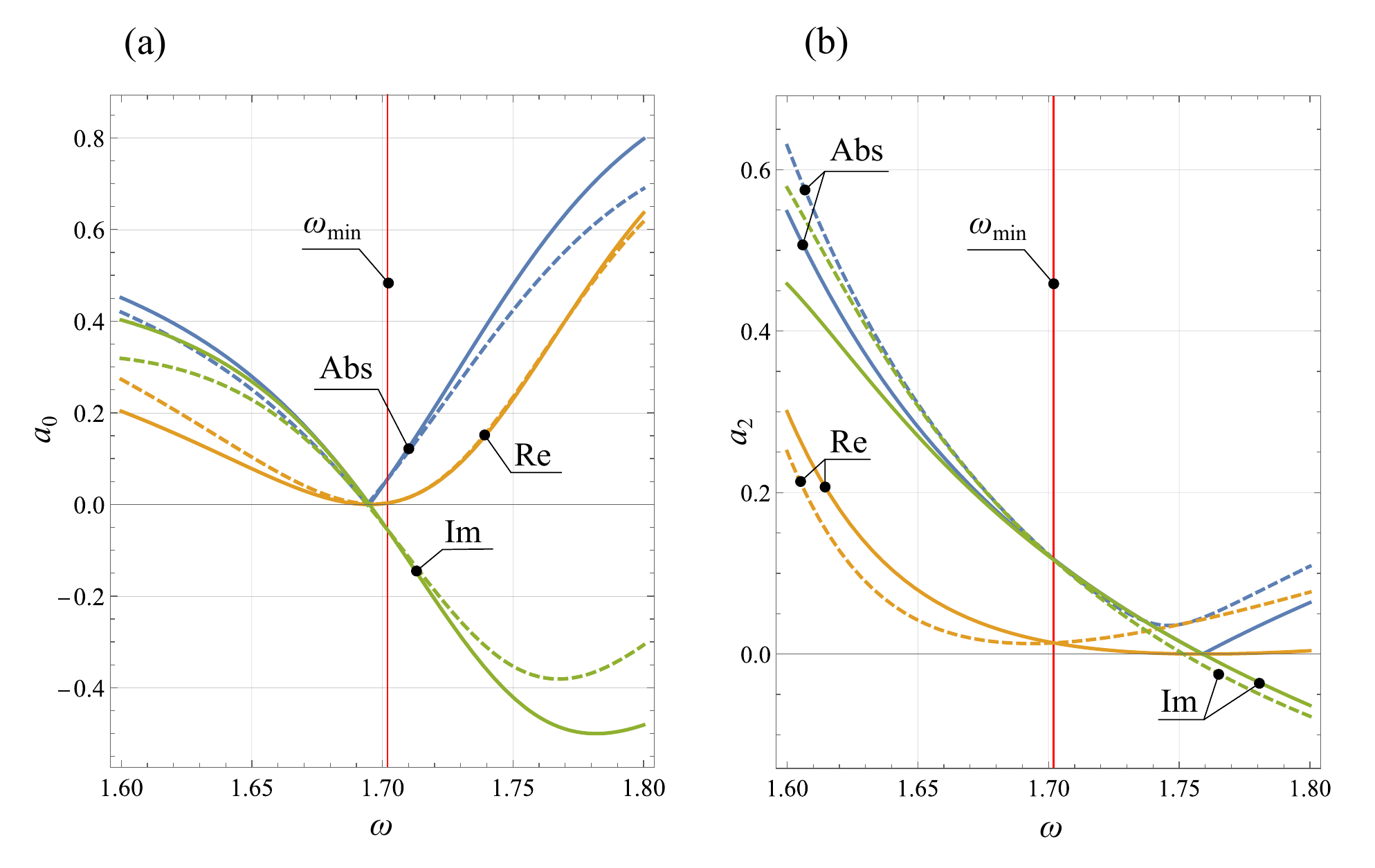}
\caption{The same as that in Fig.~\ref{fig:d} for $a$-modes. }\label{fig:a}
\end{figure}

\subsubsection{HO-model}

A steady-state solution of Eq.~(\ref{eq:driven_HO}) with the r.h.s equals $A_0\exp[-i\omega t]$ is
\begin{equation}\label{eq:f_steady}
f_s(t) = F_s(\omega)e^{-i\omega t};\;\; F_s(\omega)\equiv -\frac{|A_0|e^{i\varphi}}{\omega^2-\omega_0^2-2i\gamma\omega}.
\end{equation}
Thus, to apply the HO-model to the discussed light scattering problem, the following parameters should be fixed: the complex drive amplitude $A_0\equiv |A_0|e^{i\varphi}$, the frequency $\omega_0$ and the damping \mbox{factor $\gamma$.}
In the case of the $d$-modes the values of $\omega_0$ and $\gamma$ are given by the position of the nearest to $\omega_{\rm sim}$ complex pole of the steady-state coefficient $d_\ell(\omega)$. The remaining undefined parameters {are $A_0$ and $\varphi$. They are} readily obtained from the equality $F_s(\omega_{\rm sim}) = d_\ell(\omega_{\rm sim})$. {The results of this approximation at $\omega_{\rm sim}=\omega_{\rm min}$ and the values of the other parameters specified at the main text, are shown in Fig.~\ref{fig:d}}

The case of the PEC-modes is more tricky since neither $\omega_0$, nor $\gamma$ can be obtained in the same easy manner as that for the $d$-modes. Once again, it is convenient to present the complex amplitude $A_0$ in the form $A_0 = |A_0|e^{i\varphi}$. Then,
the maximum of $|F_s(\omega)|^2$ equals
\begin{equation}\label{eq:|A0|max}
\mathop{\rm Max}\limits_\omega\{|F_s(\omega)|^2\} \equiv |F_s|^2_{\rm max}= \frac{|A_0|^2}{4\gamma^2\omega_{0\gamma}^2},
\end{equation}
and is achieved at
\begin{equation}\label{eq:omega_max}
\omega=\omega_{\rm max}^{\rm (HO)}\equiv \sqrt{\omega^2_0-2\gamma^2}.
\end{equation}
We remind that $\omega_{0\gamma} = \sqrt{\omega_0^2-\gamma^2}=\sqrt{\omega_{\rm max}^{\rm (HO)2}+\gamma^2}$, see Eqs.~\eqref{eq:driven_HO}--\eqref{eq:foff}, \eqref{eq:omega_max}. Let us consider $\omega_{\rm max}^{\rm (HO)}$ as a new independent parameter instead of $\omega_0$.
Next, according to the procedure described in the main text, we require that $\omega_{\rm max}^{\rm (HO)} = \omega^{\rm(PEC)}_{{\rm max}(\ell)}$, where \mbox{$|a^{\rm(PEC)}_\ell(\omega^{\rm(PEC)}_{{\rm max}(\ell)})|^2_{\rm max} \equiv \left|a^{\rm(PEC)}_\ell\right|^2_{\rm max}$} is the maximal value of the corresponding quantity. This fixes the value of $\omega_{\rm max}^{\rm (HO)}$ by the shape of the profile $|a^{\rm(PEC)}_\ell(\omega)|^2$.

The condition \mbox{$|F_s|^2_{\rm max} = \left|a^{\rm(PEC)}_\ell\right|^2_{\rm max}$} expresses $|A_0|^2$ in terms of $\gamma$. To fix $\gamma$ we require that $|F_s(\omega_{\rm sim})|^2 = |a^{\rm(PEC)}_\ell(\omega_{\rm sim})|^2$, where $\omega_{\rm sim}$ is the carrier frequency of the incident pulse. It gives rise to a biquadratic equation for $\gamma$. {After some algebra, its only solution satisfying the condition Re$\,\gamma<0$, Im$\,\gamma =0$ may be presented in the following form:}
\begin{equation}\label{eq:gamma}
\gamma \! =-\frac{\omega_{\rm max}^{\rm (HO)}}{\sqrt{2}} \left\{-1 +\!\sqrt{1+\frac{\left(\omega_{\rm max}^{\rm (HO)2}-\omega_{\rm sim}^2\right)^2\left|a^{\rm(PEC)}_\ell\right|^2_{\rm sim}}{\omega_{\rm max}^{\rm (HO)4}\left(\left|a^{\rm(PEC)}_\ell\right|^2_{\rm max}\!-\!\left|a^{\rm(PEC)}_\ell\right|^2_{\rm sim}\right)}} \right\}^{1/2}
\end{equation}
where $\left|a^{\rm(PEC)}_\ell\right|^2_{\rm sim}\equiv \left|a^{\rm(PEC)}_\ell(\omega_{\rm sim})\right|^2$. {Nonnegativity of the expressions under the signs of radicals is seen straightforwardly.}

Now, the last unfixed parameter $\varphi$ is readily obtained from the condition that the phase of $a^{\rm(PEC)}_\ell(\omega_{\rm sim})$ equals the one of $F_s(\omega_{\rm sim})$. The values of $\left|a^{\rm(PEC)}_\ell\right|^2_{\rm max,\,sim}$ and $\omega=\omega^{\rm(PEC)}_{{\rm max}\,(\ell)}$ are obtained numerically according to the definition of $a^{\rm(PEC)}_\ell$, see Eq.~(\ref{eq:al-dl}).

The application of this procedure results in the values of the parameters presented in the main text, see Eqs.~(\ref{eq:PEC_OMEGA}) and (\ref{eq:PEC_An}). Note that while the approximation errors for the PEC modes is large, the final approximations for $a_\ell$ in the proximity of $\omega_{\rm sim}$ are quite accurate, cf. Figs.~\ref{fig:aPEC} and ~\ref{fig:a}.

{Next, there are some points, where $|a^{\rm(PEC)}_\ell(\omega)|$ vanishes, see, e.g., Fig~\ref{fig:aPEC}(a). If $\omega_{\rm sim}$ coincides with such a point, i.e., $\left|a^{\rm(PEC)}_\ell(\omega)\right|=0$, the phase of $a^{\rm(PEC)}_\ell(\omega)$ is indeterminate, and our method to fix the complete set of the HO-model parameters seemingly fails. However, in this case, $A_0$ in the r.h.s. of Eq.~(\ref{eq:driven_HO}) should be set to zero, the PEC-mode does not contribute to the model dynamic, and the corresponding parameters merely are not required.}

{Another point of a special interest is the one, when $\omega_{\rm sim}=\omega_{\rm max}$. In this case the straightforward application of Eq.~\eqref{eq:gamma} gives rise to an indeterminate form $0/0$. As usual, it means that we have to consider the limit $\omega_{\rm sim}\rightarrow\omega_{\rm max}$. Expanding in Eq.~\eqref{eq:gamma} $|a^{\rm(PEC)}_\ell(\omega)|^2$ about the point $\omega_{\rm sim}=\omega_{\rm max}^{\rm (HO)}$ in powers of $\delta\omega =\omega_{\rm sim}-\omega_{\rm max}^{\rm (HO)}$ we readily obtain that
\begin{equation}\label{eq:Limit_gamma}
\lim_{\omega_{\rm sim}\rightarrow\omega_{\rm max}^{\rm (HO)}}\!\!\!\! \!\!\!\!\gamma =-\frac{\omega_{\rm max}^{\rm (HO)}}{\sqrt{2}} \left\{-1+\sqrt{\frac{4}{\alpha ^2 \omega_{\rm max}^{\rm (HO)2}}+1}\right\}^{1/2},
\end{equation}
where
\begin{equation*}
\alpha^2\equiv -\frac{1}{2\left|a^{\rm(PEC)}_\ell\right|^2_{\rm max}}\left(\frac{\partial^2 \left|a^{\rm(PEC)}_\ell\right|^2_{\rm max}}{\partial\omega^2}\right)_{\!\!\omega_{\rm max}^{\rm (HO)}}\!\!\!\!\!\!>0.
\end{equation*}}

%
%\subsection{Overall dynamic}
%The comparison of the overall dynamic of the scattering obtained by the numerics~\cite{Tribelsky_Mirosh:PRA_2019} with that described by the models is presented in Figs.~\ref{fig:CM_full} and \ref{fig:HO_full}.
%\begin{figure}
%\centering
%\includegraphics[width=.85\textwidth]{Num_CM}
%\caption{$Q_{\rm sca}(t)$ obtained by the direct numeric integration of the complete set of Maxwell's equations \cite{Tribelsky_Mirosh:PRA_2019} and that described by the TCMT-model; $t=0$ corresponds to the moment, when the scattered radiation for the first time is detected by the measuring monitors. The moment $t=\tau$ is clearly seen by the abrupt change of the dynamic from almost sinusoidal to essentially non-sinusoidal. Note the overshoot of the first oscillation behind both edges of the incident pulse (i.e., behind the points $t=0$ and $t=\tau$) for the model relative to that for the numerics, explained by the instantaneous excitation of the background partition for the TCMT-model, see the discussion after Eq.~(\ref{eq:a_l^TCMT_t>tau}) in the main text. }\label{fig:CM_full}
%\end{figure}
%\nopagebreak
%\begin{figure}
%\centering
%\includegraphics[width=.85\textwidth]{Num_HO}
%\caption{The same as that in Fig.~\ref{fig:CM_full} for the HO-model. The overshoot does not occur.}\label{fig:HO_full}
%\end{figure}

%\bibliographystyle{ieeetr}
\bibliography{Nanoph}

\end{document}